\documentclass[]{rAMF2e}
\usepackage{listings}
\usepackage[center]{caption}
\usepackage{centernot}
\usepackage{url}
\usepackage{booktabs}
\usepackage{textcomp}
\usepackage[update,prepend]{epstopdf}
\DeclareMathOperator*{\rect}{rect}

\begin{document}
	\doi{}
	\issn{}  \issnp{}
	\jvol{00} \jnum{00} \jyear{2013} %\jmonth{January--March}
	\def\jobtag{}
	\publisher{Unpublished}
	\jname{}
	
	\markboth{Fabien {Le Floc'h}}{Draft}
	
	\title{Notes on the SWIFT method based on Shannon Wavelets for Option Pricing}
	\author{Fabien {Le Floc'h}}
	\affil{}
	\date{\today}
	\received{v0.1 released January 2018}

\maketitle
\begin{abstract}
This note shows that the cosine expansion based on the Vieta formula is equivalent to a discretization of the Parseval identity. We then evaluate the use of simple direct algorithms to compute the Shannon coefficients for the payoff. Finally, we explore the efficiency of a Filon quadrature instead of the Vieta formula for the coefficients related to the probability density function.
\end{abstract}

\section{Introduction}
 \citet{ortiz2016highly} describe a novel approach to the pricing of European options under models with a known characteristic function, based on Shannon Wavelets, referred to as the SWIFT method hereafter. 
This note shows that the cosine expansion based on Vieta's formula is equivalent to a discretization of Parseval's identity. We then evaluate the use of simple direct algorithms to compute the Shannon coefficients for the payoff. Finally, we explore the efficiency of a Filon quadrature instead of Vieta's formula for the coefficients related to the probability density function.

The equivalence with Parseval's identity is also stated in \citep{maree2017pricing}.

\section{Equivalence with Parseval's identity}
With the SWIFT method, the price at time $t$ of a Vanilla Put option of maturity $T$ and log-moneyness $x=\ln\frac{F}{K}$, with $K$ the strike and forward $F$ is
\begin{equation}
v(x,t) = B(t,T) \sum_{k=k_1}^{k_2} c_{m,k} V_{m,k}
\end{equation}
where 
\begin{align}
	c_{m,k} &= \left\langle f | \phi_{m,k} \right\rangle = 2^{\frac{m}{2}}\int_{\mathbb{R}} f(x) \phi\left(2^m x - k\right) dx,,\\
 	V_{m,k} &= \int_{I_m} v(y,T)\phi_{m,k}(y)dy\,,\label{eqn:vmk}\\
	\phi_{m,k}(x) &= 2^{\frac{m}{2}}\phi\left(2^m x - k\right)\,,\\
	\phi(x) &= \frac{\sin \pi x}{\pi x}
\end{align}
and $k_1,k_2,m \in \mathbb{Z}$, $m >= 1$ suitably chosen, $f$ the probability density function and $v(y,T)$ is the payoff at maturity with $y=\ln\frac{F(T,T)}{K}$, that is  $v(y,T)=K|1-e^y|^+$ for a vanilla Put option.

In \citep{ortiz2016highly}, the coefficients $c_{m,k}$ and $V_{m,k}$ are computed using an approximation based on Vieta formula for the cardinal sinus:
\begin{equation}
\phi(x) \approx \frac{1}{2^{J-1}}\sum_{j=1}^{2^{J-1}}\cos\left(\frac{2j-1}{2^J}\pi x\right)
\end{equation}
where $J$ is chosen sufficiently large.

As mentioned in paragraph 3.1.2 of their paper, $c_{m,k}$ can also be computed by  Parseval's identity:
\begin{equation}
c_{m,k} =  \left\langle f | \phi_{m,k} \right\rangle = \frac{1}{2\pi} \left\langle \hat{f} | \hat{\phi}_{m,k} \right\rangle
\end{equation}
where $\hat{f},  \hat{\phi}_{m,k}$ are the Fourier transforms of $f$ and $\phi_{m,k}$. In particular $\hat{f}(z) = \psi(-z)$ where $\psi$ is the characteristic function and
\begin{equation}
\hat{\phi}_{m,k}(w)=\frac{e^{-i \frac{k}{2^m}w}}{2^{\frac{m}{2}}} \rect\left(\frac{w}{2^{m+1}\pi}\right)
\end{equation} where $\rect$ is the rectangular function, that is $\rect(x) = 1$ for $|x| < \frac{1}{2}$, $\rect(x)=\frac{1}{2}$ for $|x|=\frac{1}{2}$, $\rect(x) = 0$ for $|x| > \frac{1}{2}$.

%We thus have
%\begin{equation}
%c_{m,k} = \frac{1}{2^{m/2}2\pi} \int_{-2^m \pi}^{2^m \pi} \hat{f}(w) e^{i k/2^m w}dw
%\end{equation}
Via a the change of variable $t=\frac{w}{2^{m+1}\pi}$, we obtain
\begin{align}
c_{m,k} &= 2^{\frac{m}{2}}\int_{-\frac{1}{2}}^{\frac{1}{2}} \left[\hat{f}(2^{m+1}\pi t)e^{i 2\pi k t} \right]dt\\
&=2^{\frac{m}{2}+1}\Re\left[\int_{0}^{\frac{1}{2}} \hat{f}(2^{m+1}\pi t)e^{i 2\pi k t} dt\right]\label{eqn:cmk_parseval}
\end{align}
as $\Re(\psi(x)) = \frac{ \psi(x) + \overline{\psi(x)}}{2} =  \frac{ \psi(x) + \psi(-x)}{2}$.

Let us now discretize in $2^{J-1}$ equidistant steps equation (\ref{eqn:cmk_parseval}) of size $\frac{1}{2^{J}}$ at the mid-points $t_j = \frac{j-\frac{1}{2}}{2^{J}} $ for $j=1,2,...,2^{J-1}$, we obtain
\begin{align}
c_{m,k}^\star &= \frac{2^{\frac{m}{2}}}{2^{J-1}}\sum_{j=1}^{2^{J-1}} \Re\left[\hat{f}(2^{m}\pi 2 t_j)e^{i \pi k 2t_j} \right]\\
&= \frac{2^{\frac{m}{2}}}{2^{J-1}}\Re\left[\sum_{j=1}^{2^{J-1}} \hat{f}\left( \frac{2^{m}\pi (2j-1)}{2^{J}} \right) e^{2i \pi k \frac{j-\frac{1}{2}}{2^{J}}} \right]\\
&= \frac{2^{\frac{m}{2}}}{2^{J-1}} \Re\left[ e^{i \pi \frac{k}{2^J}} \sum_{j=0}^{2^{J-1}-1}\hat{f}\left( \frac{2^{m}\pi (2j-1)}{2^{J}} \right) e^{2 i \pi k \frac{j}{2^J}} \right]
\label{eqn:cmk_parseval_discrete}
	\end{align}
This is exactly equation (24) of \citep[p. B127]{ortiz2016highly} which corresponds to their expansion based on Vieta's formula. Their expansion is thus equivalent to the mid-point quadrature applied to Parseval's identity.

A particularly important property of equation \ref{eqn:cmk_parseval_discrete} is that it can be computed by fast Fourier transform (FFT). The typical FFT algorithm computes the tranform (or inverse transform) from index $0$ to $n$. Here, we start with a negative index $k_1$. The coefficients can be obtained with the relation 
\begin{align}
	\mathbf{c} &= \frac{2^{\frac{m}{2}}}{2^{J-1}} \Re\left[ \mathbf{e}^T \mathcal{F}^{-1}\left\{ \mathbf{f} \right\} \right]
\end{align}
where $\mathcal{F}^{-1}$ is the unscaled inverse discrete Fourier transform of size $2^J$, the vector $\mathbf{f}$ has elements $f_j= \hat{f}\left( \frac{2^{m}\pi (2j+1)}{2^{J}} \right) e^{2i\pi \frac{k_1 j }{2^J}}$ and the vector $\mathbf{e}$ has elements $e_l= e^{i\pi \frac{ l+k_1 }{2^J}}$. We also assumed that $\hat{f}\left( \frac{2^{m}\pi (2j+1)}{2^{J}} \right) = 0$ for $j\geq 2^{J-1}$.
This leads to a very efficient way to compute the coefficients $c_{m,k}$, for all $k$, together. In practice, this means that the bounds $k_1 \leq k < k_2$ are chosen so that $k_2 - k_1 < 2^J$.

In particular, if we center the interval around zero, that is for $k_1 = -2^{J-1}$, we can save a bit of computation by directly using  $f_j= \hat{f}\left( \frac{2^{m}\pi (2j+1)}{2^{J}} \right)$ and swapping $(g_{0},...,g_{2^{J-1}})$ with $(g_{2^J-1},...g_{2^J-1})$ where $\mathbf{g} = \mathcal{F}^{-1}\left\{ \mathbf{f} \right\}$.

\section{Alternative quadratures}
\subsection{Trapezoidal}
Instead of the mid-point method, we could have considered the trapezoidal method, this would result in
\begin{equation}
c_{m,k}^\star = \frac{2^{\frac{m}{2}}}{2^{J-1}} \Re\left[ \sum_{j=0}^{2^{J-1}-1}w_j\hat{f}\left( \frac{2^{m}\pi (2j)}{2^{J}} \right) e^{2 i \pi k \frac{j}{2^J}} \right]
\end{equation}
where $w_j=1$ for $j\geq 1$ and $w_0=\frac{1}{2}$.

The fast inverse discrete  Fourier transform  of length $2^J$ can be directly used to compute $c_{m,k}$ by using  $f_j= \hat{f}\left( \frac{2^{m}\pi (2j)}{2^{J}} \right)$ for $ 1 \leq j < 2^{J-1}$, $f_j = 0$ for $2^{J-1} \leq j$, and $f_0= \frac{1}{2}\hat{f}(0)$.

We will see in the numerical examples that it can be much more accurate than the mid-point method.

In the same framework, we could also explore other quadratures, such as the Simpson's quadrature. The problem is that those tend to behave worse than the midpoint or trapezoidal rules on oscillatory functions. In fact, the trapezoidal rule can achieve exponential convergence on oscillatory functions \citep{johnson2011numerical,trefethen2014exponentially}. In the case of the probability density transform function $\hat{f}$, this can be also be seen from the Euler-Maclaurin formula where as all the derivatives $\hat{f}^{(2l+1)}\left(2^m \pi\right)$ will be small if the characteristic function decreases exponentially.

% 3/8. We don't require the interval to be divided in $n$ parts with $n$ divisible by three as  $\mathbf{f}$ is padded with zeros. In this case the weights are
%$w_{3j+1} = \frac{9}{8}$, $w_{3j+2} = \frac{9}{8}$, $w_{3j+3} = \frac{6}{8}$ for $j=0,..,\frac{2^J}{3}$ and $w_0=\frac{3}{8}$.
%
%$f(x_j)$ with $x_j = 2^m \pi 2j / 2^J$, $x_{j+1} -x_j = 2^m \pi / 2^{J-1}$

\subsection{Adaptive Filon}
Instead of quadrature with a fixed number of steps, we can use an adaptive Filon quadrature to compute the coefficients $c_{m,k}$ by equation (\ref{eqn:cmk_parseval}). This is particularly interesting since the cost of computing the characteristic function is relatively high.

Is it more important to reduce its number of evaluations than to use Fast Fourier Transform tricks to compute $c_{m,k}$?
We will explore this in the numerical examples (section \ref{sec:numerical_examples}).

An alternative adaptive quadrature, close in spirit, is to use an adaptive cubic-Hermite quadrature to integrate $\hat{f}$, and use the integration nodes to compute the piecewise cubic Hermite interpolant of $\hat{f}$. Then we can use the trapezoidal-FFT approach on a dense discretization. This saves explicit computations of the characteristic function while still allowing the use of the FFT algorithm.

\section{Sine and Exponential integrals for the payoff}
For a Vanilla Put option, the payoff at maturity is $V(y,T)= K(1-e^y)^+$. According to equation (\ref{eqn:vmk}), the payoffs coefficients are then
\begin{align}
	V_{m,k} &=  K2^{\frac{m}{2}} \int_{a}^0 (1-e^y) \frac{\sin \left(\pi \left(2^m y -k\right)\right)}{\pi \left(2^m y -k\right)}dy\\
	&=   K2^{\frac{m}{2}} \int_{a}^0 \frac{\sin \left(\pi \left(2^m y -k\right)\right)}{\pi \left(2^m y -k\right)}dy -   K2^{\frac{m}{2}} \int_{a}^0 e^y \frac{\sin \left(\pi \left(2^m y -k\right)\right)}{\pi \left(2^m y -k\right)}dy\\
	&= \frac{K}{2^{\frac{m}{2}}\pi} \int_{\pi \left(2^m a -k\right)}^{-\pi k}\frac{\sin t}{t}dt - \frac{K e^{\frac{k}{ 2^m}}}{2^{\frac{m}{2}}\pi} \int_{\pi \left(2^m a -k\right)}^{-\pi k}e^{\frac{t}{\pi 2^m}}\frac{\sin t}{t}dt
\end{align}
The first integral corresponds the sine integral $\textsf{Si}(x)=\int_{0}^{x} \frac{\sin t}{t} dt$. Many efficient algorithms exist to compute it \citep{macleod1996rational,jin1996computation}. Most mathematical software (for example Octave, Matlab) or libraries (for example netlib) include the function.  It can effectively be considered as a closed form function.

The second integral can be reduced to evaluations of the complementary exponential integral $\textsf{Ein}(z)= \int_0^z \frac{1 - e^{-t}}{t}dt$ in the complex plane. In deed, it can be verified that we have the identity
\begin{equation}
\int_{0}^1 \frac{e^{-a t} \sin (b t)}{t} dt = \Im \textsf{Ein}(a+ib)
\end{equation}
The complementary exponential integral is related to the exponential integral $\textsf{Ei}(z)=-\int_z^{\infty} \frac{e^{-t}}{t}dt$ by the relation $\textsf{Ein}(z)=\gamma + \ln |z| + i\Im(-z)|\arg(-z)| -\textsf{Ei}(-z)
$. Again many efficient algorithms exist to compute the complementary exponential integral \citep{amos1990algorithms,jin1996computation,pegoraro2011evaluation}. 

In terms of those special functions, the coefficients are:
\begin{equation}
V_{m,k} = \frac{K}{2^{\frac{m}{2}}\pi} e^{\frac{k}{ 2^m}} \Im\left[\textsf{Ein}\left(-\frac{t_a}{\pi 2^m}+i t_a\right)-\textsf{Ein}\left(-\frac{t_0}{\pi 2^m}+i t_0\right)\right] - \frac{K}{2^{\frac{m}{2}}\pi} \left[\textsf{Si}(t_a) - \textsf{Si}(t_0)\right]
\end{equation}
with $t_a=\pi \left(2^m a -k\right)$ and $t_0=-\pi k$.

The expansion based on Vieta's formula might require thousands of terms to reach an acceptable accuracy (Table \ref{tbl:vieta_accuracy}). With the same number of terms, a Simpson 3/8 quadrature is more accurate and faster to compute. Our simple implementation of the algorithm from \citet{pegoraro2011evaluation} is much faster and achieves machine epsilon accuracy while the algorithm from the CERN libary Mathlib \citep{kolbig1990exponential} is even faster for a close to machine epsilon accuracy as it relies on simple rational and pad\'e expansions in the zone of interest. In practice, the implementation of the SWIFT method will still benefit from a cache table of $V_{m,k}$ for example for $m \in \{2,...,8\}$ and $k\in \{-512,...,512\}$.

\begin{table}[h]
	\centering{\caption{$V_{m,k}$ for $m=6$, $k=-1$, $a=-1$. Vieta's formula or Simpson's quadrature use $2^{J-1}$ terms. \label{tbl:vieta_accuracy}}
	\begin{tabular}{l r r}\toprule
		Method & Value & Time(ns) \\\midrule
		Vieta $J=5$ & -0.0555195115435162 & 600 \\
Simpson $J=5$ & -0.0020905045216672& 520\\
		Vieta $J=10$ &  0.0020428901436639& 17300 \\
		Simpson $J=10$ & 0.0020420973936057& 15300\\
		CERN &    0.0020420954069492& 420\\
		Pegoraro & 0.0020420954069488& 2500\\\bottomrule
	\end{tabular} 
}
\end{table}
	%2t_j = 1/2 * 2j+1 /2^{J-1}
%	#1. vieta cos expansion is the same as Parseval + simple quadrature (rectangular/trapezoidal) discretization
%	#1.2 Vmk slow convergence: J>=10 often necessary in my tests. Unfortunate that for J fixed the basis is not orthonormal.
%	#2. Both cmk and Vmk can be computed directly via Parseval
%	#3. Filon could provide better convergence
%	#4. Exp integral function algorithms exist in many libraries. Faster Vmk?
%	## Error not so simple to control: interplay between m and k1/k2.

	\section{Alternative payoff coefficients}
The interval $[a, b]$ is centered along the spot $F(0,T)$, we can express the payoff in terms of the spot $F(0,T)$ instead of the strike $K$. This leads to
	\begin{align}
		V_{m,k} &= 2^{\frac{m}{2}} \int_{a}^b F\left|\frac{K}{F} - e^y\right|^+ \frac{\sin \left(\pi \left(2^m y -k\right)\right)}{\pi \left(2^m y -k\right)}dy\\
		 &= F2^{\frac{m}{2}} \int_{a}^z \left(e^z - e^y\right) \frac{\sin \left(\pi \left(2^m y -k\right)\right)}{\pi \left(2^m y -k\right)}dy\\
		 &= Ke^{-z} 2^{\frac{m}{2}} \int_{a}^z \left(e^z - e^y\right) \frac{\sin \left(\pi \left(2^m y -k\right)\right)}{\pi \left(2^m y -k\right)}dy\label{eqn:swift_payoff_new}
	\end{align}
	where $z=\ln \frac{K}{F}$ and $y=\ln \frac{S_T}{F}$.
	
	In terms of those special functions, the coefficients are:
	\begin{equation}
	V_{m,k}(z) = \frac{Ke^{\frac{k}{ 2^m}-z}}{2^{\frac{m}{2}}\pi} \Im\left[\textsf{Ein}\left(-\frac{t_a}{\pi 2^m}+i t_a\right)-\textsf{Ein}\left(-\frac{t_z}{\pi 2^m}+i t_z\right)\right] - \frac{K}{2^{\frac{m}{2}}\pi} \left[\textsf{Si}(t_a) - \textsf{Si}(t_z)\right]
	\end{equation}
	with $t_a=\pi \left(2^m a -k\right)$ and $t_z=\pi \left(2^m z -k\right)$.

The price of the option of strike $K$ corresponds then to $v(0,t)$. The coefficients $c_{m,k}$ become independent of $x$. This is nearly equivalent to the Levy based equation (33) in \citep{ortiz2016highly} that defines the coefficients $V_{m,k}^{\alpha}(x)$. The difference lies in the interval considered. In their paper, $V_{m,k}^{\alpha}(x) = \int_{a}^b K|1-e^u|^+ \phi_{m,k}(u+z) du$ with $u=\ln \frac{S_T}{K}$. This can be rewritten as 
\begin{equation}\label{eqn:swift_payoff_classic}
V_{m,k}^{\alpha}(x) = K \int_{a+z}^{z} \left(1 - e^{t-z}\right) \phi_{m,k}(t)dt
\end{equation}
 with the change of variable $t=u+z$. The interval $[a, b]$ is thus shifted from $z$ upwards. Our choice of interval is more accurate as it corresponds directly to the Levy characteristic function, while their interval is based on the shifted Levy characteristic function. Also their Levy formulation (as well as ours) leads to options prices different from the classic formulation: for the two to be equivalent, the integers $k_1$ and $k_2$ should be adjusted to $k_1 = 2^m (a+z)$ and $k_2 = 2^m (b+z)$. But then some of the density coefficients need to be recomputed at each strike as the window $[k_1, k_2]$ moves forward as $z$ increases and the Levy approach loses in efficiency.
 	
Another advantage of having $c_{m,k}$ independent of the strike is that the integers $k_1$ and $k_2$ can also be determined in a strike independent manner from the value of the density coefficients $c_{m,k}$ and the area under the curve defined by the probability density (which should sum to one minus a user-defined tolerance) as explained by \citet{ortiz2016highly}, instead of relying on the relatively rough guess given by the cumulants (the fixed interval $[a, b]$). With the cumulants approach, it is not always obvious how large the truncation level $L$ should be chosen to achieve a desired accuracy.
 	\section{Alternative FFT-compatible payoff coefficients}
 	In a similar fashion to \citet{maree2017pricing}, we start from the definition 
 	\begin{equation}
 	\frac{\sin(\pi x)}{\pi x} = \int_{0}^{1} \cos(\pi xw)dw\,.
 	\end{equation}
 	We can then choose an appropriate discretization that has good convergence, and is allows computation of the payoff coefficients $V_{m,k}$ by the FFT. The choice from \citet{ortiz2016highly} is equivalent to the mid-point quadrature. On this problem, the Trapezoidal rule would not lead to an increase in accuracy\footnote{It can be shown that the mid-point is actually more accurate by a factor of two.}. A particularly simple an effective choice is the second Euler-Maclaurin summation formula, that is the Euler-Maclaurin extension to the mid-point rule. 
 		\begin{equation}
 	\frac{\sin(\pi x)}{\pi x} \approx \frac{1}{N} \sum_{n=0}^{N-1} \cos \left(\pi x \frac{2n+1}{2N}\right) + \frac{\pi x}{24 N^2} \sin (\pi x)\,.
 	\end{equation}
 	Using the above in equation (\ref{eqn:swift_payoff_new}) leads to
 	\begin{align}
 		V_{m,k} &= 	\frac{Ke^{-z} 2^{\frac{m}{2}}}{N} \sum_{n=1}^{N-1} \int_{a}^z \left(e^z - e^y\right) \cos \left(\pi \left(2^m y -k\right) \frac{2n+1}{2N}\right) dy\nonumber \\
 		&+\frac{\pi }{24 N^2}Ke^{-z} 2^{\frac{m}{2}}  \int_{a}^z \left(2^m y -k\right)\left(e^z - e^y\right) \sin \left(\pi \left(2^m y -k\right) \right) dy
 	\end{align}
 	
% 	We then apply the Trapezoidal rule with Euler-Maclaurin first derivative correction
% 	 	\begin{equation}
% 	\frac{\sin(\pi x)}{\pi x} \approx \frac{1}{N}\left[\frac{1}{2} + \sum_{n=1}^{N-1} \cos \left(\pi x \frac{n}{N}\right) + \frac{1}{2}\cos \left(\pi x \right) +\frac{\pi x}{12 N} \sin (\pi x) \right]\,.
% 	\end{equation}
% 	Using the above in equation (\ref{eqn:swift_payoff_new}) leads to
% 	\begin{align}
% 	V_{m,k} &= 	 \frac{Ke^{-z} 2^{\frac{m}{2}}}{2N} \int_{a}^z \left(e^z - e^y\right)  dy\nonumber\\
% 	&+	 \frac{Ke^{-z} 2^{\frac{m}{2}}}{N} \sum_{n=1}^{N-1} \int_{a}^z \left(e^z - e^y\right) \cos \left(\pi \left(2^m y -k\right) \frac{n}{N}\right) dy\nonumber \\
% 	&+ \frac{Ke^{-z} 2^{\frac{m}{2}}}{2N} \int_{a}^z \left(e^z - e^y\right) \cos \left(\pi \left(2^m y -k\right) \right) dy\nonumber\\
% 	  &+\frac{\pi }{12 N^2}Ke^{-z} 2^{\frac{m}{2}}  \int_{a}^z \left(2^m y -k\right)\left(e^z - e^y\right) \sin \left(\pi \left(2^m y -k\right) \right) dy
% 	\end{align}
 	Let $C_n(a,z) =  \int_{a}^z \left(e^z - e^y\right) \cos \left(\pi 2^m y \frac{n}{N} \right) dy$ and $S_n(a,z) = \int_{a}^z \left(e^z - e^y\right) \sin \left(\pi 2^m y \frac{n}{N} \right) dy$. Using the trigonometric cos and sin identities, we obtain
 		\begin{align}
 		V_{m,k} &=\frac{Ke^{-z} 2^{\frac{m}{2}}}{N}\sum_{n=0}^{N-1}C_{n+\frac{1}{2}}(a,z) \cos\left(\pi k\frac{2n+1}{2N}\right) + S_{n+\frac{1}{2}}(a,z) \sin\left(\pi k\frac{2n+1}{2N}\right)\nonumber\\
 		&- \frac{(-1)^k \pi k}{24 N^2}Ke^{-z} 2^{\frac{m}{2}} S_N(a,z)\ - \frac{ (-1)^k}{24 N^2}  Ke^{-z} 2^{\frac{m}{2}}  D(a,z)
 	\end{align}
 	with
 	\begin{align}
 		D(a,z)&=  e^{z}\frac{\left( \left( {{{{p}_{m}}}^{4}}+{{{{p}_{m}}}^{2}}\right) z-3{{{{p}_{m}}}^{2}}-1\right) \,\sin{\left( {{p}_{m}}z\right) }+\left( \left( {{{{p}_{m}}}^{3}}+{{p}_{m}}\right) z+2{{{{p}_{m}}}^{3}}\right)\,\cos{\left( {{p}_{m}}z\right) }}{{p}_{m}(p_m^4+2{{{{p}_{m}}}^{2}}+1)}\nonumber\\
 		&+ e^{z}\frac{\left( {{{{p}_{m}}}^{4}}+2{{{{p}_{m}}}^{2}}+1\right) \sin{\left( a\,{{p}_{m}}\right) }+\left( -a\,{{{{p}_{m}}}^{5}}-2a\,{{{{p}_{m}}}^{3}}-a\,{{p}_{m}}\right) \cos{\left( a\,{{p}_{m}}\right) }}{{p}_{m}(p_m^4+2{{{{p}_{m}}}^{2}}+1)}\nonumber\\
 		&+e^a\frac{\left( \left( -a-1\right) \,{{{{p}_{m}}}^{4}}+\left( 1-a\right) \,{{{{p}_{m}}}^{2}}\right) \sin{\left( a\,{{p}_{m}}\right) }+\left( a\,{{{{p}_{m}}}^{5}}+\left( a-2\right) \,{{{{p}_{m}}}^{3}}\right) \cos{\left( a\,{{p}_{m}}\right) }}{{p}_{m}(p_m^4+2{{{{p}_{m}}}^{2}}+1)}\,,\\
 		C_n(a,z) &= e^z \frac{\sin\left(q_{n,m} z\right) - q_{n,m} \cos\left(q_{n,m} z\right)}{q_{n,m}(1+q_{n,m}^2)}-e^z\frac{\sin\left(q_{n,m} a\right)}{q_{n,m} }\nonumber\\
 		&+e^a \frac{\cos\left(q_{n,m} a\right)+q_{n,m}\sin\left(q_{n,m} a\right) }{1+q_{n,m}^2}\,,\\
 		S_n(a,z) &= -e^z \frac{\cos\left(q_{n,m} z\right) + q_{n,m} \sin\left(q_{n,m} z\right)}{q_{n,m}(1+q_{n,m}^2)}+e^z\frac{\cos\left(q_{n,m} a\right)}{q_{n,m} }\nonumber\\
 		&+e^a \frac{\sin\left(q_{n,m} a\right)-q_{n,m}\cos\left(q_{n,m} a\right) }{1+q_{n,m}^2}\,,\\
 		q_{n,m} &= \frac{n}{N}p_m\,,\\
 		p_m &= \pi 2^m\,.
 	\end{align}
 In particular, $S_N$ and $D$ are independent of $k$. Computing $V_{m,k}$ with the Euler-Maclauring correction for all $k$ requires only $k$ more multiplications than the mid-point quadrature. The sum over $n$ corresponds to the mid-point quadrature, and can be computed with two fast Fourier transforms of size $N$ (see appendix A).
	\section{Choice of $m$ and $k_1, k_2$}
	The SWIFT method accuracy is fully determined by the choice of the scale $m$ and the truncation $k_1,k_2$. There is some interplay between those since the scale $m$ also determines the truncation of the characteristic function: the characteristic function will not be evaluated beyond $2^m \pi$.

	If we want to use the radix-2 FFT algorithm to compute the payoff coefficients $V_{m,k}$, there is little reason not to use $k_2-k_1=2^J$, centered on zero, where a reasonably good guess for $J$ can be obtained from the model characteristic function cumulants. In the evaluation of a single option strike, the cost of computing the payoff coefficients will dominate the cost of evaluating the price based on the sum of the $V_{m,k}$ multiplied by the (precomputed) density coefficients $c_{m,k}$. Furthermore, the number of coefficients must be a power of two and must include $[k_1,k_2)$.
	
	The scale $m$ is more challenging to guess. It can be guessed from the rule used to truncate the integral of the more standard Fourier based approach from \citet{andersen2010interest}, refined in \cite{floc2013fourier}. It then directly depends on the asymptotic behaviour of the characteristic function. \citet{maree2017pricing} propose a simple iterative method to determine $n$ automatically (with very few iterations on $m$).
	
	\section{Numerical examples}\label{sec:numerical_examples}
\subsection{Payoff coefficients $V_{m,k}$ and the FFT}
Vieta's formula is not very efficient to compute a single coefficient  $V_{m,k}$  but as we compute close to $2^J$ coefficients the FFT improves its performance significantly. For $2^10$ coefficients, Vieta's formula end up around six times faster than the CERN algorithm.
\begin{table}[h]
	\centering{\caption{Time in microseconds taken to compute $V_{m,k}$ for $k=0,...,2^{J-1}-1$ with $m=6$, $a=-1$.}
		\begin{tabular}{r r r}\toprule
			$J$ & FFT & CERN \\\midrule
			5 & 1.7 & 10.7\\
			10 & 56 & 360 \\\bottomrule
		\end{tabular} 
	}
\end{table}

While the raw difference in performance is impressive. It is more interesting to look at the actual performance difference when pricing vanilla Put options under the Heston stochastic volatility model. We consider two different Heston parameter sets for two distinct option maturities. This leads to two vastly different truncation ranges $[a,b]$, computed according to the Heston cumulants. As a result $2^J=2^{12}=k_2-k_1$ for the first set and $2^J=2^8 = k_2-k_1$ for the second set. Ignoring the initialization time where the $c_{m,k}$ are computed, which needs to be done only once per option expiry, the direct CERN algorithm is between five to eight times slower.

\begin{table}[h]
	\centering{\caption{Heston parameter sets.\label{tbl:heston_params}}
	\begin{tabular}{l r r r r r r r}\toprule	 
		Name & $v_0$ & $\kappa$ &  $\theta$ &$\sigma$ & $\rho$ & $F$ & $T$ \\\midrule
		Set 1 & 0.1 & 1.0 & 0.1 & 1.0 & -0.9 & 1.0 & 2 days\\ 
		Set 2 & 0.0225 & 0.1 & 0.01 &,2.0 & 0.5 & 1000000 & 1 year\\ \bottomrule
	\end{tabular} 
}
\end{table}

\begin{table}[h]
	\centering{\caption{Time in milliseconds taken to compute the Put option price under two different Heston parameter sets.}
		\begin{tabular}{l l r r r}\toprule
			Heston & Method & Price & Error & Time (ms) \\\midrule
			Set 1 (J=12) & FFT & 117.9149 &  -1.4704 & 0.250\\
			  & CERN & 117.9144 &  -1.4708 & 1.370\\
			Set 2 (J=8) & FFT &  0.006361& -3.49e-15 & 0.016 \\
			  & CERN &  0.006361& 6.42e-13 & 0.101 \\\bottomrule
			%1.0064 CosF-128 0.006360762834222444 1.5370789710478228e-11 5.464µs swift can use m=7 instead of m=8 and  we are down to 9 vs 7 micros. => not really slower
		\end{tabular} 
	}
\end{table}

\subsection{New payoff coefficients versus the original formulation}
We consider options of maturity 2 days (short) in order to make the issue more visible and we consider the Heston parameters s $\kappa=1.0, \theta=0.1, \sigma=1.0, \rho=-0.9, v_0=0.1$, along with a forward price at valuation time $F=1.0$. Those parameters are not extreme, and are in the typical range of a Heston fits to market option prices.

\begin{figure}[htbp]
	\caption{\label{fig:swift_truncation}Error in Vanilla option prices of maturity 2 days with Heston parameters $\kappa=1.0, \theta=0.1, \sigma=1.0, \rho=-0.9, v_0=0.1, F=1.0$ using a truncation levels $L=12$ and scale $m=8$.}
	\begin{center}
		\includegraphics[width=\textwidth]{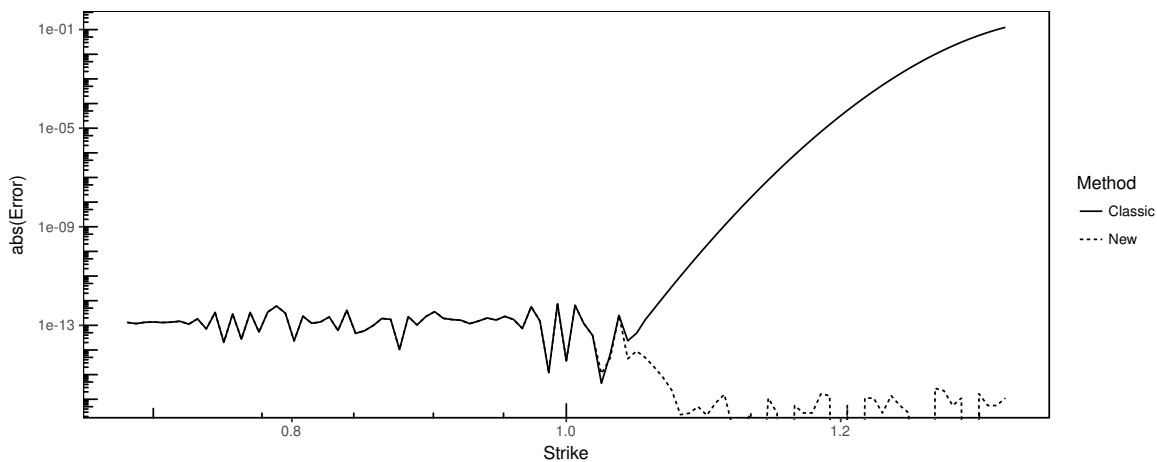}
	\end{center}
\end{figure}

In Figure \ref{fig:swift_truncation}, we look at the absolute error in price for a scale $m=8$ and a truncation $L=12$ based on the Heston cumulants. This truncation corresponds to an interval $[a,b]=[-0.2815, 0.2810]$. Our reference is the price obtained by the Lord-Kahl optimal alpha method \citet{lord2007optimal}. We consider two ways of computing the payoff coefficients: the classic payoff formula of \cite{ortiz2016highly} represented by equation (\ref{eqn:swift_payoff_classic}), and our new formula represented by equation (\ref{eqn:swift_payoff_new}). We make sure that the density coefficients are computed with maximum accuracy by using a large $J$, so that the overall error is dominated by error in the payoff formula.
We stop at strike $K=1.32$ since then $\ln \frac{K}{F} >b$. Figure \ref{fig:swift_truncation} shows that the error of the new formula stays below $10^{-13}$, close to machine epsilon while the error of the classic formula can be as high as $1.5\cdot 10^{-2}$ when the strike approaches the upper boundary $F e^b$.

\subsection{Density coefficients $c_{m,k}$ and quadratures}
We consider the same Heston model parameters as in the previous section. The trapezoidal rule is three to six times more accurate than the mid-point rule (or equivalently the formula from \citet{ortiz2016highly} based Vieta's formula) across strikes and on both Heston sets. Both rules use exactly the same number of points.
\begin{table}[h]
	\centering{\caption{Price of an out-of-the-money option under two different Heston parameter sets.}
		\begin{tabular}{l l r r r r}\toprule
			Heston  & Method & Strike & Price & Error  \\\midrule
			Set 1 ($m=8, J=12$) & Midpoint & 250000 & 114.51 & -4.87 \\
			& Trapezoidal & 250000 &  117.91 & -1.47 \\
			& Midpoint & 4000000 & 3866.59 & -85.33 \\
			& Trapezoidal & 4000000 &  3931.09 & -20.82 \\
			Set 2 ($m=6, J=5$) & Midpoint & 1.0064 & 0.0063611& 3.97e-07 \\
			& Trapezoidal &  1.0064 & 0.0063606& -7.39e-08 \\
			 & Midpoint & 1.064 &  4.77e-06 & 5.09e-07 \\
			& Trapezoidal &  1.064 &  4.18e-06 & -8.22e-08 \\
			\bottomrule
			%1.0064 CosF-128 0.006360762834222444 1.5370789710478228e-11 5.464µs swift can use m=7 instead of m=8 and  we are down to 9 vs 7 micros. => not really slower
		\end{tabular} 
	}
\end{table}

We now look at the time to initialize the SWIFT method for a given option maturity. This corresponds to the calculation of the density coefficients $c_{m,k}$, either with the FFT applied on the trapezoidal quadrature, or with the direct adaptive Filon quadrature on a relative tolerance of $10^{-8}$ (which leads to a similar accuracy as the FFT approach).
\begin{table}[h]
	\centering{\caption{Initialization time of the SWIFT method for two different Heston parameter sets and different quadratures.}
		\begin{tabular}{l l r r r}\toprule
			Heston & Method & Points & Time (microseconds)\\\midrule
Set 1 & FFT & 4096  & 433 \\
& Filon  & 585 & 76000 \\
Set 2 & FFT & 32  & 16 \\
&Filon  & 497 & 588 \\\bottomrule
		\end{tabular} 
}
\end{table}
For a similar accuracy, the initialization based on the adaptive Filon quadrature is slower by a factor of more than 32 although the characteristic function is evaluated 585 times compared to 2048 times for the FFT calculation. There is then a lot of room if we were to make the FFT density calculation adaptive by doubling successively the interval $[k_1,k_2]$.%to reach  time.

\section{Conclusion}
The use of the fast Fourier transform (FFT) to compute the payoff coefficients is particularly important and makes the SWIFT method competitive with some of the fastest pricing methods such as  COS method of \citet{fang2008novel}. Our alternative formula centered on the forward is more accurate in general than the original payoff coefficients formula from \citet{ortiz2016highly} while being of equivalent computational cost.

The calculation of the density coefficients also benefits from the FFT, even though the related characteristic function is relatively expensive to compute. The FFT based on the trapezoidal rule is much more accurate than the original formula from \citet{ortiz2016highly} for a slightly lower computational cost.  Using more fancy adaptive quadratures is no so useful.
A simple adaptive scheme based successively doubling the truncation interval $[k_1,k_2]$ according to the accuracy of the area underneath the curve is good enough.

\bibliographystyle{rAMF}
\bibliography{swift_notes}
\appendix
\section{Computing the discrete Cosine and Sine transforms together from the FFT}
The calculation of the $V_{m,k}$ by the formula described in Appendix A of \citet{ortiz2016highly} is the sum of a type 2 discrete cosine transform (DCT) and a type 2 discrete sine transform (DST). It can be summarized by the following equation
\begin{align}
	V_{m,k} &= \sum_{j=0}^{N-1} a_j \cos\left(\pi k\frac{j+\frac{1}{2}}{N}\right) + b_j \sin\left(\pi k\frac{j+\frac{1}{2}}{N}\right)
\end{align}
with $N= 2^{\bar{J}-1}$ for some positive integer $\bar{J}$.
\citet{makhoul1980fast} gives a simple algorithm to compute the DCT of size $N$ with one FFT of size $N$.
We simply initialize the FFT coefficients $c_j$ with:
\begin{equation}
	c_j = a_{2j} \quad\,,\quad	c_{N-1-j} = a_{2j+1} \quad \textmd{ for } j = 0,..., \frac{N}{2}-1
\end{equation}
and then from the result of the FFT $\hat{c}$, the DCT coefficients $\hat{a}$ are
\begin{equation}
\hat{a}_k = \Re\left[ \hat{c}_j e^{-i \pi\frac{k}{2N}} \right]
\end{equation}

Makhoul does not specify the equivalent formula for the DST, but we can do something similar. We first initialize the FFT coefficients $c_j$ with:
\begin{equation}
c_j = b_{2j} \quad\,,\quad	c_{N-1-j} = -b_{2j+1} \quad \textmd{ for } j = 0,..., \frac{N}{2}-1
\end{equation}
and then from the result of the FFT $\hat{c}$, the DST coefficients $\hat{b}$ are
\begin{equation}
\hat{b}_k = -\Im\left[ \hat{c}_j e^{-i \pi\frac{k}{2N}} \right]
\end{equation}

For maximum performance, the two FFTs can reuse the same sine and cosine tables. And the last step of the DCT and DST can be combined together.
\end{document}